\documentclass[aps,showpacs,twocolumn]{revtex4}

\usepackage{graphicx}

\begin{document}

\title{Antiferromagnetically coupled CoFeB/Ru/CoFeB trilayers}

\author{N. Wiese}
\email[Electronic Mail: ]{mail@nilswiese.de}

\affiliation{Siemens AG, CT MM 1, Paul-Gossen-Str. 100, 90592
Erlangen, Germany}
\affiliation{University of Bielefeld, Nano Device Group,
 Universit\"atsstr.25, 33615 Bielefeld, Germany}
%\homepage[]{Your web page}
%\thanks{}

\author{T. Dimopoulos}
\affiliation{Siemens AG, CT MM 1, Paul-Gossen-Str. 100, 90592
Erlangen, Germany}

\author{M. R\"uhrig}
\affiliation{Siemens AG, CT MM 1, Paul-Gossen-Str. 100, 90592
Erlangen, Germany}

\author{J. Wecker}
\affiliation{Siemens AG, CT MM 1, Paul-Gossen-Str. 100, 90592
Erlangen, Germany}

\author{H. Br\"uckl}
\affiliation{University of Bielefeld, Nano Device Group,
Universit\"atsstr.25, 33615 Bielefeld, Germany}

\author{G. Reiss}
\affiliation{University of Bielefeld, Nano Device Group,
Universit\"atsstr.25, 33615 Bielefeld, Germany}

\date{\today}

\begin{abstract}
This work reports on the magnetic interlayer coupling between two amorphous CoFeB layers, separated by a thin Ru spacer. We observe an antiferromagnetic coupling which oscillates as a function of the Ru thickness $x$, with the second antiferromagnetic maximum found for $x=1.0$ to 1.1 nm. We have studied the switching of a CoFeB/Ru/CoFeB trilayer for a Ru thickness of 1.1 nm and found that the coercivity depends on the net magnetic moment, i.e. the thickness difference of the two CoFeB layers. The antiferromagnetic coupling is almost independent on the annealing temperatures up to $300^{\circ}$C while an annealing at $350^{\circ}$C reduces the coupling and increases the coercivity, indicating the onset of crystallization. Used as a soft electrode in a magnetic tunnel junction, a high tunneling magnetoresistance of about $50\%$, a well defined plateau and a rectangular switching behavior is achieved.
\end{abstract}

\pacs{75.47.Np, 75.50.Kj}
\maketitle

Spin valves composed of magnetic tunnel junctions (MTJ) have gained considerable interest in recent years due to their high potential as sensor elements in applications as read heads \cite{Ho01}, angle \cite{Berg99} or strain sensors \cite{Loehndorf02a} and as programmable resistance in data storage (MRAM) \cite{Gallagher97} or even data processing \cite{Richter02a}.

The basic design of a spin valve consists of a hard magnetic reference electrode separated from the soft magnetic sense or storage layer by a tunnel barrier like Al$_2$O$_3$. The reference layer usually is an artificial ferrimagnet (AFi) exchange biased by a natural antiferromagnet, in which the AFi consists of two ferromagnetic layers antiparallely coupled through a thin non-magnetic spacer. For the soft electrode, mostly NiFe has been used \cite{Koch98} and only recently this has been substituted by amorphous alloys of 3d ferromagnets with metalloids like B or Si, showing not only low switching fields \cite{Kaeufler02} but also high tunnel magnetoresistance (TMR) \cite{Wang04, Tsunekawa04}.

The concept of an artificial ferrimagnet allows one to further adjust the magnetic properties of the soft layer. Compared with a single ferromagnetic layer the AFi can be regarded as a rigid magnetic body with a reduced magnetic moment and enhanced anisotropy. The gain in coercivity, $H_{\mathbf c}^{\mathbf{AFi}}$, can be expressed by a factor $Q$:
\begin{eqnarray}
H_{\mathbf c}^{\mathbf {AFi}} = Q \cdot H_{\mathbf c}^{\mathbf {SL}} \quad \mbox{with} \quad Q = \frac{M_1 t_1 + M_2 t_2}{M_1 t_1 - M_2 t_2} \label{Qvalue}
\end{eqnarray}
where $M_1, M_2$ and $t_1, t_2$ are the saturation magnetization and the thickness of the two composite ferromagnetic layers and $H_{\mathbf c}^{\mathbf {SL}}$ is the single layer coercivity \cite{Berg96}. The $Q$ value and thus $H_{\mathbf c}^{\mathbf {AFi}}$ can be easily tailored by modifying the thickness of the ferromagnetic layers. Such AFi free layers show a smaller switching field distribution \cite{Sousa02} and patterned elements with small aspect ratio more easily retain a single domain structure \cite{Tezuka03a}. In addition, demagnetizing fields at the edges of sub-$\mu$m sized elements can be reduced.

While so far only AFi systems of polycrystalline materials like CoFe and NiFe have been used, it was the purpose of this study to extend the knowledge about AFi soft electrodes towards amorphous alloys. As the amorphous alloy we chose Co$_{60}$Fe$_{20}$B$_{20}$ because of its high spin polarization, leading to high TMR values \cite{Wang04}, while still using a Ru spacer. All samples investigated have been deposited by magnetron sputtering on thermally oxidized SiO$_2$ wafers at a base pressure of $5 \cdot 10^{-8}$ mbar. A magnetic field of approximately 4 kA/m was applied during deposition in order to induce the easy axis in the magnetic layers. The AFi was grown on a 1.2 nm thick Al layer, oxidized in an Ar/O${_2}$ plasma for 0.8 min without breaking the vacuum, to have similar growth conditions as in a MTJ.

\begin{figure}
    \includegraphics[width=8.6cm]{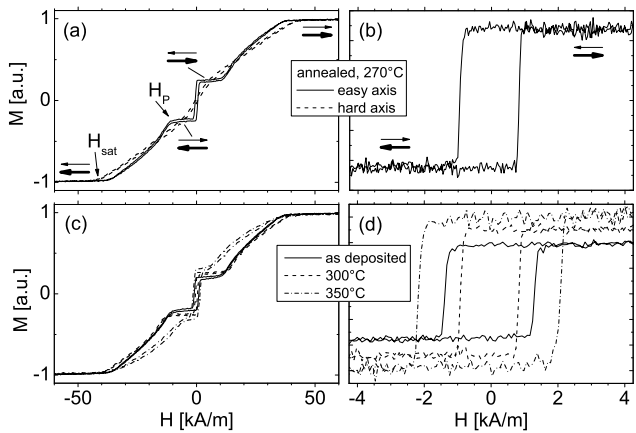}
    \caption{Magnetization loops of the system Ta(5) / Al(1.2, oxid.) / CoFeB(4) / Ru(1.1) / CoFeB(3) / Ta(10) at easy and hard axis (a)/(b) and at different annealing temperatures (c)/(d).} \label{fig:agmloops}
\end{figure}

In order to investigate the coupling phenomena of the CoFeB-based AFi and the behavior of the coercivity with varying net moment, two series of samples have been prepared. Series A consists of CoFeB(4) / Ru($x$) / CoFeB(3), where the thickness of the nonmagnetic Ru-spacer has been varied between $x=0.7$ and 1.2 nm in steps of 0.1 nm. Series B consists of CoFeB(t$_1$) / Ru(1.1) / CoFeB(3) where the thickness of the ferromagnetic layer t$_1$ in contact with the Al$_2$O$_3$ layer is varied between 3.5 and 5 nm, which gives nominal $Q$-values between 13 and 4, respectively. Additional two single CoFeB layers with thicknesses of 3.5 nm and 6 nm have been deposited for comparison. All samples were capped with a Ta layer to protect the multilayers from oxidation.

To study the temperature stability of the coupling, the samples have been annealed on a hot plate at constant temperatures between $200$ and $350^{\circ}$C for 15 min and have been protected from oxidation by a constant Ar-flow. The ramp-down times varied between 1 and 2 hours, depending on the applied annealing temperature. A field of approximately 400 kA/m was
applied along the deposition-induced easy axis during annealing and cooling down.

\begin{figure}
  \includegraphics[width=6.5cm]{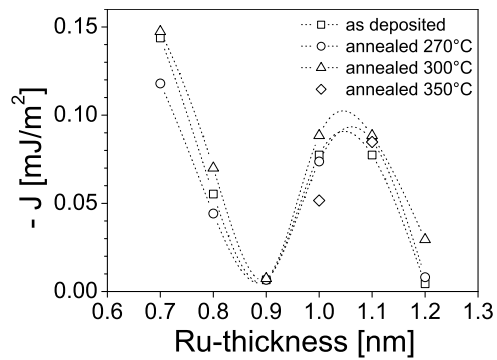}
  \caption{(a) Coupling $J$ vs. Ru-spacer thickness for the samples of series A taken from AGM measurements and evaluated for different annealing temperatures. (b) Dependence of the coercive field $H_{\mathbf c}^{\mathbf {AFi}}$ of the AFi extracted from minor loops on the measured $Q$-value of the AFi of series B. The line is a fit confirming the linear behavior.} \label{fig:coupling_coercivity}
\end{figure}

A typical room temperature magnetization curve, $M(H)$, of an antiferromagnetically (AF) coupled system with a Ru spacer thickness of 1.1nm is shown in figure \ref{fig:agmloops}(a) after annealing at $270^{\circ}$C. The $M(H)$ curve shows a well defined anisotropy and a good antiparallel alignment during the magnetization reversal of the net moment of the AFi. From the hard axis loop one can extract the saturation field, $H_{\mathbf {sat}}$. Using $M_{1,2} = 860 \pm 50$ kA/m, as determined from alternating gradient magnetometer (AGM) measurement of the single CoFeB layer with 6nm thickness we calculate the coupling energy ${J = - \mu_0 H_{\mathbf {sat}} \frac{M_1 t_1 M_2 t_2}{M_1 t_1 +M_2 t_2}}$ \cite{Berg97}. The coupling $J$ of the samples of series A shows an oscillating behavior in the antiferromagnetic region. The antiferromagnetic coupling has a maximum of $-0.08$ mJ/m$^2$ at spacer thicknesses of 1.0 to 1.1 nm [figure \ref{fig:coupling_coercivity} (a)] and therefore it is approximately by a factor of 10 lower than in artificial ferrimagnets of CoFe/Ru/CoFe \cite{Berg99,Dimopoulos}. The coupling does not change significantly with annealing up to temperatures of 300 to $350^{\circ}$C \cite{Wiese04b}.

Besides, all samples show a drastic increase of coercivity after an anneal at $350^{\circ}$C [e.g. shown for a system of series B in figure \ref{fig:agmloops}(d)]. This abrupt increase is observed regardless of the thickness of the CoFeB and the materials interfacing it. This strongly suggests that the origin of the coercivity increase is not caused by interdiffusion, but the change from amorphous to polycrystalline phase as also reported for other CoFeB-alloys \cite{Jimbo97}. A detailed study of the structural properties is currently under investigation. For 350$^\circ$C the AF coupling in the AFi is still maintained, excluding the possibility of pinhole formation at this temperature [see figure \ref{fig:agmloops}(c)].

For the investigation of the dependence of coercivity on the $Q$-value of the AFi, samples of series B have been studied. By extracting the saturation moment $M_{\mathbf s}=M_1 + M_2$ and the net moment $M_{\mathbf {net}}=M_1 - M_2$ of the AFi from the AGM measurements, we can calculate the measured $Q$-value, defined as ${Q_{\mathbf{meas}}=M_{\mathbf s}/M_{\mathbf{net}}}$, and the individual moments of the layers $M_1$,$M_2$. It is found for all samples that $Q_{\mathbf {meas}}$ depends on the annealing temperature and is significant lower than the nominal $Q$-value. This discrepancy in $Q$-value can be explained by the observed thicker magnetically dead-layers of the upper CoFeB-layer in comparison to the bottom layer. Since the samples are well protected from oxidation, as confirmed by Auger depth profiling, this indicates a stronger intermixing of the upper CoFeB interfaces, that leads to a increase in $M_{\mathbf {net}}$ and therefore a decrease in $Q_{\mathbf{meas}}$.

Plotting $H_{\mathbf c}$ as a function of $Q_{\mathbf {meas}}$ we clearly see the expected linear behavior [figure \ref{fig:coupling_coercivity} (b)]. The linear fit of the data gives $H_{\mathbf c}^{\mathbf{AFi}}=(0.29 \cdot Q - 0.17)$ kA/m, leading to a single layer coercivity of $0.29$ kA/m, in good agreement with the measured coercivity of $H_{\mathbf c}^{\mathbf {SL}}=0.27$ to 0.35 kA/m for the 3.5 nm thick single layer.

\begin{figure}
    \includegraphics[width=8.6cm]{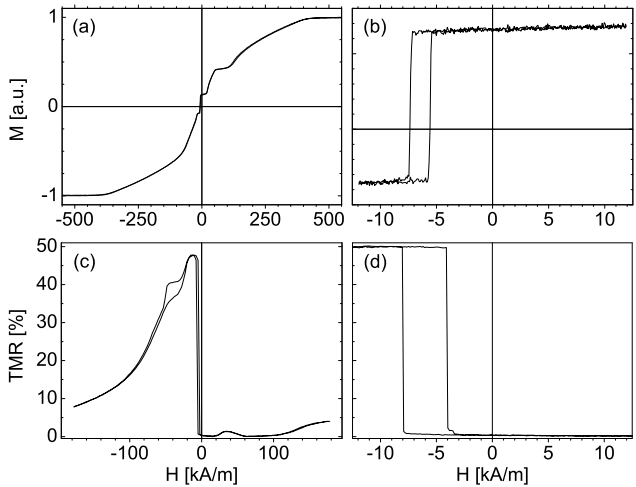}
    \caption{(a)+(b) Magnetization major and minor loop, (c)+(d) magnetoresistive measurement for the complete TMR multilayer stack system (sample C) after a field annealing procedure at $270^{\circ}$C.} \label{fig:fullstack}
\end{figure}

Finally, we have integrated an amorphous CoFeB-AFi as a soft-magnetic electrode in a complete MTJ stack system. As the layer sequence IrMn(8) / Co$_{90}$Fe$_{10}$(3) / Ru(0.8) / Co$_{90}$Fe$_{10}$(3) / Al(1.2, oxid. 0.8 min) / CoFeB(3) / Ru(1.1) / CoFeB(4) on a appropriate seed-layer was chosen and the exchange bias was set with a field anneal at $270^{\circ}$C. Figure \ref{fig:fullstack}(a)/(b) shows the magnetization loop after a field anneal at $270^{\circ}$C. The soft magnetic AFi sense layer has a coercivity of $H_{\mathbf c}=1$ kA/m that is comparable to the same AFi investigated in series B, indicating only a weak influence of the additional underlayers on the coercivity. A well defined plateau field ($H_P \approx 17$ kA/m) and a saturation field of $H_{\mathbf {sat}}=36$ kA/m is observed for the top AFi. From this values the coupling energy can be obtained, which is comparable with the corresponding AFi of series B. The high N\'eel-coupling of $H_{\mathbf N} = 6.4$ kA/m most likely results from the roughness induced by the buffer layer which has not been optimized. The magnetoresistance measurements [figures \ref{fig:fullstack} (c)/(d)] show a high TMR amplitude of $\sim 50 \%$, comparable to results with a soft electrode of amorphous CoFeB single layers presented elsewhere \cite{Dimopoulos_JAP04}, and a rectangular switching of the soft electrode AFi at $H_{\mathbf c}=1.9$ kA/m. The increase in coercivity is probably caused by pinning at the boundary edge \cite{Meyners03}, since SEM images show very high edge roughness of the measured elements with $12.5 \times 12.5$ $\mu$m$^2$ and the reduction of the offset is probably related to demagnetizing fields.

The investigation of magnetic interlayer coupling in systems consisting of two ferromagnetic layers of amorphous CoFeB separated by a thin Ru spacer has been presented. Oscillating antiferromagnetic coupling, well known for crystalline 3d alloys, has been observed, which is stable against annealing up to temperatures of $300-350^{\circ}$C. Used as an artificial ferrimagnet (AFi) with the Ru thickness at the second antiferromagnetic maximum, a clear linear dependence of the coercivity on the $Q$ value has been found. This allows one to tailor the coercivity in a wide range by changing the net magnetic moment of the AFi. Used as the free layer in a MTJ a well defined switching with high TMR values up to $50 \%$ has been achieved.

The authors wish to thank J. Bangert and G. Gieres for fruitful discussions, and H. Mai for experimental support.

\end{document}